\documentclass[aps,pra,preprint,groupedaddress,superscriptaddress,showpacs]{revtex4-1}
\usepackage{CJK}
\usepackage{amsmath}
\usepackage{longtable}
\usepackage[dvipdfm,  %pdflatex,pdftex这里决定运行文件的方式不同
            pdfstartview=FitH,
            CJKbookmarks=true,
            bookmarksnumbered=true,
            bookmarksopen=true,
            colorlinks, %注释掉此项则交叉引用为彩色边框(将colorlinks和pdfborder 同时注释掉)
            pdfborder=001,   %注释掉此项则交叉引用为彩色边框
            linkcolor=blue,
            anchorcolor=blue,
            citecolor=blue
            ]{hyperref}
\usepackage{mathrsfs}
\usepackage{multirow}
\usepackage{tabularx}
\usepackage{array}
\usepackage{graphicx}
\usepackage{bm}
\begin{document}

%\preprint{}

\title{Exploration of the magnetic-field-induced $5s5p$~$^3P_0$~-~$5s^2$~$^1S_0$ forbidden transition in bosonic Sr atom}


\author{Benquan Lu}
\affiliation{National Time Service Center, 710000 Xi'an, China}
\affiliation{The University of Chinese Academy of Sciences, 100088 Beijing, China}
\affiliation{Institute of Applied Physics and Computational Mathematics, 100088 Beijing, China}
\author{Yebing Wang}
\affiliation{National Time Service Center, 710000 Xi'an, China}
\affiliation{The University of Chinese Academy of Sciences, 100088 Beijing, China}
\author{Jianxin Han}
\affiliation{National Time Service Center, 710000 Xi'an, China}
\affiliation{The University of Chinese Academy of Sciences, 100088 Beijing, China}
\author{Shougang Zhang}
\affiliation{National Time Service Center, 710000 Xi'an, China}
\author{Jiguang Li}
\thanks{Li\_jiguang@iapcm.ac.cn}
\affiliation{Institute of Applied Physics and Computational Mathematics, 100088 Beijing, China}
\author{Hong Chang}
\thanks{changhong@ntsc.ac.cn}
\affiliation{National Time Service Center, 710000 Xi'an, China}
\noaffiliation


\date{\today}

\begin{abstract}
We presented experimental and theoretical studies of the effect of an external magnetic field on the forbidden transition in the bosonic Sr atom. In our ultra-cold atomic system, the excitation fraction of $5s5p$~$^3P_0$~-~$5s^2$~$^1S_0$ forbidden transition was measured under the circumstance of the different magnetic field strengths by using the normalized detection method. Based on perturbation theory, we calculated the magnetic-field-induced $5s5p$~$^3P_0$~-~$5s^2$~$^1S_0$ transition rate.
The excitation fraction as a function of the magnetic field strength was deduced according to the calculated results. A good agreement was found between the experimental measurements and the calculations. This study should be helpful in evaluating the magnetic field effects on the forbidden transition rate with higher accuracy. Moreover, it can help to understand the ultra-cold atomic interaction in the external magnetic field.
\end{abstract}

%\keywords{}

\maketitle

\section{INTRODUCTION}
The forbidden lines are commonly used for studying astrophysical and laboratory plasmas \cite{PhysRevLett.97.183001}. The forbidden lines are also very important in optical clocks as their narrow natural linewidth can be used for laser cooling and trapping experiments \cite{NJP13125010,NJP17055008} and clock transition interrogation \cite{NJP18113002,NJP16073023}. Moreover, in the ultra-cold atomic system, one can use the forbidden line to measure the lifetime of the metastable state \cite{PhysRevLett.92.153004}, to observe the motion-dependent nonlinear dispersion \cite{PhysRevLett.114.093002}, and to probe the many-body interaction \cite{PhysRevLett.95.223002,PhysRevLett.90.063002}. However, it is a challenge to predict and determine the rate of the forbidden transition due to its sensitivity to the electron correlations and the relativistic effects \cite{PhysRevA.93.032506,Phy.Scr.89.114002}.

The $5s5p$~$^3P_0$~-~$5s^2$~$^1S_0$ transition in neutral Sr atom, as a typical $E$1 forbidden transition, has been extensively studied in experiments and theories for its potential applications in quantum computing \cite{PhysRevLett.95.060502}, optical atomic clocks \cite{Tino2013,RevModPhys.87.637,Nature2011,PhysRevA.81.023402}, and atom interferometers \cite{PhysRevA.92.053619}. This forbidden transition is induced by the hyperfine interaction in the fermionic atom, and thus referred to as the hyperfine-induced transition. In contrast to fermionic Sr, bosonic Sr has a simpler level structure and higher natural abundance. The $5s5p$~$^3P_0$~-~$5s^2$~$^1S_0$ transition can be induced as well, but by the external magnetic field instead of the internal magnetic field --- the hyperfine interaction. In the bosonic Sr optical clock, this clock transition rate depends on the magnetic field strength \cite{PhysRevLett.96.083001,Baillard:07}. However, the forbidden transition rate has not been determined yet.

In this work, we measured the excitation fraction of $5s5p$~$^3P_0$~-~$5s^2$~$^1S_0$ forbidden transition using the normalized detection method under the circumstance of the different magnetic field strengths in our $^{88}$Sr ultra-cold atomic system. Meanwhile, we carried out a calculation on the magnetic-field-induced $5s5p$~$^3P_0$~-~$5s^2$~$^1S_0$ transition rate in the framework of the multi-configuration Dirac-Hartree-Fock (MCDHF) method. Based on this, we deduced the excitation fraction as a function of the magnetic field strength. A comparison between the experimental results and theoretical calculations was made and we found that the theoretical calculations are in good agreement with our measurements.
\section{Experimental measurement of $5s5p$~$^3P_0$~-~$5s^2$~$^1S_0$ excitation fraction}
\subsection{Apparatus}
\begin{figure}
 \centering
 \includegraphics[width=0.45\textwidth]{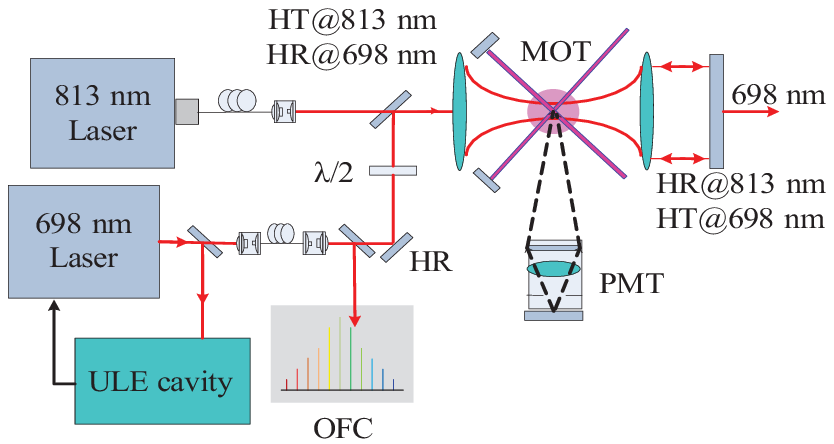}
 \includegraphics[width=0.4\textwidth]{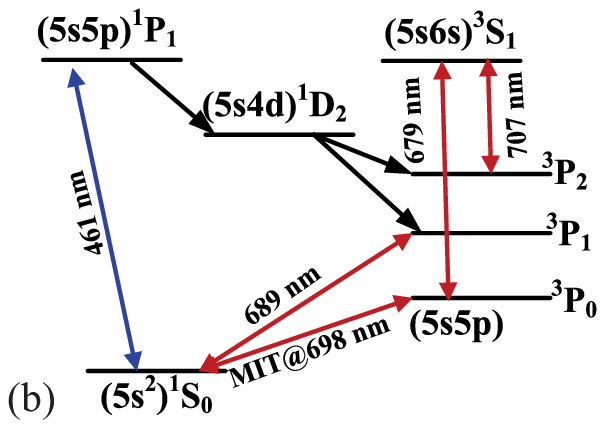}
 \caption{\label{3} (a) Experimental setup for the forbidden transition. HWP: $\lambda$/2 wave plate; DM: dichroic mirror;
 OFC: optical frequency comb; PMT: photomultiplier tube; ULE: ultra-low expansion. (b) Simplified energy levels of $^{88}$Sr. The
 $^1S_0$~-~$^{1,3}P_1$ transitions were used for laser cooling and trapping. Atoms in the $^3P_{0,2}$ metastable states were transferred to $^3S_1$ state by the 679-nm and 707-nm lasers and then decayed to the ground state through $^3S_1$~$\rightarrow$~$^3P_1$~$\rightarrow$~$^1S_0$ as the spontaneous decay rate from $^3P_1$ state is much larger than that from $^3P_{0,2}$ states. The $^3P_0$~-~$^1S_0$ transition is the magnetic-field-induced transition.}
\end{figure}

Our experimental setup and the simplified energy levels of $^{88}$Sr atom were shown in Fig.~\hyperref[3]{1(a)} and \hyperref[3]{(b)}, individually. Cold atoms preparation was described in detail in Ref.~\cite{COL.30.100201}. The $^1S_0$~-~$^{1,3}P_1$ transitions were used for first and second stage magnetic-optically trapping which cooled the atoms down to a few $\mu$K. The 679-nm and 707-nm lasers were the re-pumping lasers. Atoms in the $^3P_{0,2}$ metastable states were pumped to the $^3S_1$ state by using the re-pumping lasers to drive the $^3P_0$~-~$^3S_1$ and $^3P_2$~-~$^3S_1$ transitions. Eventually, these atoms would decay to the ground state through $^3S_1$~$\rightarrow$~$^3P_1$~$\rightarrow$~$^1S_0$ as the spontaneous decay rate from $^3P_1$ state is much larger than that from $^3P_{0,2}$ states. After the two-stage cooling, the ultra-cold bosonic Sr atoms were loaded into an optical lattice which was operated at the ``magic wavelength" of $\lambda_L$ = 813~nm \cite{PhysRevA.81.023402,PhysRevA.75.020501}. The 813-nm laser beam was transferred by a single-mode-polarization-maintaining fiber. The 698-nm probe laser was locked to a stable ultra-low expansion (ULE) cavity which was placed on an enclosed vibration isolation platform, and was transferred by a noise cancelled fiber link. The line-width of the laser was about 1~Hz and the Allan deviation was 1 $\times$ 10$^{-15}$ at 1~s. This laser was split into two beams, one going to the femtosecond optical frequency comb (OFC, Menlo FC1500) to have its frequency measured. The second one passed through a $\lambda$/2 wave plate (HWP) and a mechanical shutter and combined with the lattice laser using a dichroic mirror (DM). The 813-nm laser and the 698-nm laser passed through a Glan-Taylor polarizer to make sure their polarizations, and focused by an achromatic lens (f = 200~mm) into the vacuum chamber and onto the cold atoms. The wave-vector $\bf{k}$ of the two lasers are perpendicular to the gravity. The waist of the 813-nm laser was 38~$\mu$m (1/$e^2$ radius of intensity), and the 698-nm laser beam was approximately three times larger. After exciting the vacuum chamber, the lattice laser was retro-reflected by another achromatic lens and a mirror to overlapped with the incident laser beam and formed the standing wave. The retro-reflecting mirror was low-reflecting at the 698-nm wavelength. The focusing lens were mounted on two three-dimensional translation stages respectively to optimize the optical lattice and the $^1S_0$~-~$^3P_1$ magneto-optical trap overlapping. The static magnetic field was produced by Helmholtz coils and its direction was parallel to the linear polarization of the 698-nm probe laser.
\subsection{Method}
About $10^5$ atoms were trapped in the one-dimensional optical lattice. The temperature of these atoms was 8.4~$\mu$K and the lifetime of the atoms trapped in the optical lattice was 500~ms. The $5s5p$~$^3P_0$~-~$5s^2$~$^1S_0$ forbidden transition was probed in the Lamb-Dicke regime along the lattice longitudinal axis. During the experiment, the 813-nm laser was maintained open. We used the normalized detection method \cite{refId0} to measure the $5s5p$~$^3P_0$~-~$5s^2$~$^1S_0$ excitation fraction. Firstly, we used the 698-nm probe laser pulse ($\pi$-pulse) to pump some of these atoms, namely $N_1$, to the $^3P_0$ state. The static magnetic field was also applied. Secondly, the remaining atoms, namely $N_2$, were pushed out of the optical lattice with a 461-nm detection laser pulse. Thirdly, the atoms remained in the $^3P_0$ state were transferred to the ground state by the 679-nm and 707-nm laser pulses. Finally, the pumped atoms ($N_1$) were also pumped to the $^1P_1$ state. The values of $N_1$ and $N_2$ were measured by probing the 461-nm fluorescence intensity. As the lifetime of $^3P_0$ state is very long, we neglected the population decay of $^3P_0$ state atoms in the optical lattice after the probe and detection pulses we had chosen. We assumed that atoms in the $^3P_0$ state were all transferred to the ground state. The excitation fraction was deduced from these two measurements by
\begin{equation}
P = \frac {N_1}{N_1 + N_2}.
\end{equation}

\begin{figure}
 \includegraphics{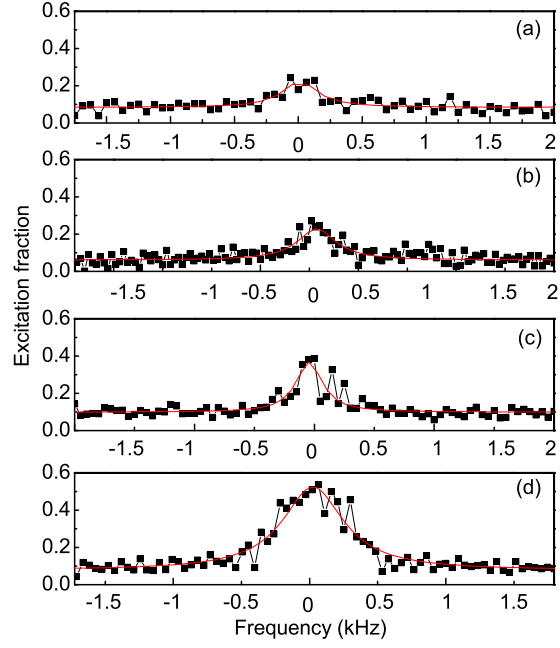}
 \caption{\label{4}Spectrum of the magnetic-field-induced $5s5p$~$^3P_0$~-~$5s^2$~$^1S_0$ transition with the magnetic field strength B = 1.0, 1.2, 1.7, 2.2~mT from (a) to (d). The red curves are Lorentzian fit of the data.}
\end{figure}

We used a photomultiplier tube (PMT) to measure the 461-nm fluorescence intensity for evaluating the numbers of the atoms in the ground state. In order to obtain a high signal-to-noise ratio of the fluorescence intensity, a 461-nm filter, a lens group and an aperture were used to eliminate the background noise. The measured spectrums were presented in Fig.~\hyperref[4]{2(a)-(d)} under the circumstance of the magnetic field strength B = 1.0, 1.2, 1.7, 2.2~mT, respectively. During the measurements, the power of the 698-nm laser was maintained constant. The spectrums were generated by repeating the detection cycle and stepping the frequency of the 698-nm laser using an acousto-optic modulator. The experimental data were fitted with Lorentzian function. From this figure, we can see that the intensity of the spectrum peak decreased and the line-width shrinked with the magnetic field strength decreasing. When B = 1.0~mT, the line-width is 180~Hz which is larger than the Fourier limited width (20~Hz) due to 45~ms interrogation pulse. The line-width is broadened by many factors, for example, the atomic collisions. Since in this paper the intensity of the spectrum peak is the focus of the discussion, the line-width broadening is neglected.
\section{magnetic-field-induced transition}
\subsection{Theory}
\subsubsection{Magnetic-field-induced transition rate}
In the presence of an external magnetic field $\textbf{B}$, the atomic Hamiltonian is \cite{PhysRevA.88.013416}
\begin{equation}
H = H_{fs} + H_{m},
\end{equation}
where $H_{fs}$ is the relativistic fine-structure Hamiltonian which include the Breit interaction and the main part quantum electrodynamical (QED) effects and $H_m$ is the Hamiltonian for the interaction between the external magnetic field and the atom. If the magnetic field does not vary throughout the atomic system, the interaction Hamiltonian $H_m$ is expressed as \cite{Hfszeeman}
\begin{equation}
H_m = (\bm{N}^{(1)} + \Delta \bm{N}^{(1)}) \cdot \textbf{\textrm{B}}.
\end{equation}
Here, $\Delta\textbf{\textit{N}}^{(1)}$ is the so-called Schwinger QED correction \cite{Hfszeeman}. For an $N$-electron atom, these two operators are
\begin{equation}
\bm{N}^{(1)} = \sum\limits_{j = 1}^N {\bm{n}^{(1)}}(j) =  \sum\limits_{j = 1}^N {{-i}\frac{\sqrt2}{2\alpha}} {r_j} (\bm{\alpha}_j \bm{C}^{(1)}(j))^{(1)},
\end{equation}
\begin{equation}
\Delta \bm{N}^{(1)} = \sum\limits_{j = 1}^N \Delta {\bm{n}^{(1)}}(j) =  \sum\limits_{j = 1}^N {\frac{{{g_s} - 2}}{2}} {\beta _j}{\bm{\Sigma}_j}.
\end{equation}
Here, $i$ is the imaginary unit, $r_j$ is the radial coordinate of the $j^{th}$ electron, $\textbf{\textit{C}}^{(1)}(j)$ is the spherical tensor operator of rank 1, $\bm{\Sigma}$$_j$ is the relativistic spin-matrix, $g_s$ is the $g$-factor of the electron spin corrected by the QED effects, $\alpha$ is the fine-structure constant and $\bm{\alpha}$ and $\beta$ are the Dirac matrices.

If we choose the direction of the magnetic field as the quantization axis, only the magnetic quantum number $M_J$ remains the good quantum number. The atomic states with the same magnetic quantum number and parity are mixed due to the interaction between the external magnetic field and the atom \cite{PhysRevA.88.013416}. Therefore, the atomic state wave function $\left|M_J \right\rangle$ can be written as
\begin{equation}
\left| M_J \right\rangle  = \sum\limits_{\Gamma J} {{d_{\Gamma J}}\left| {\Gamma JM_J} \right\rangle },
\end{equation}
where the atomic state wave functions $\left| {\Gamma JM_J} \right\rangle$ are eigenstates of the Hamiltonian $H_{fs}$, $J$ and $M_J$ are the total and magnetic quantum numbers. According to the first-order perturbation theory, the expansion coefficient $d_{\Gamma J}$ is given by
\begin{equation}
{d_{\Gamma J}} = \frac{\left\langle {\Gamma JM_J} \right| H_m \left| {{\Gamma}_0 J_0 M_0} \right\rangle}{E({{\Gamma}_0 J_0 M_0}) - E({\Gamma JM_J})},
\end{equation}
where $\left| {{\Gamma}_0 J_0 M_0} \right\rangle$ is the reference atomic state.

Therefore, the magnetic-field-induced transition (MIT) rate can be obtained by
\begin{equation}
{A_{MIT}} = \frac{{2.02613 \times {{10}^{18}}}}{{{\lambda ^3}}}{\sum\limits_q {\left| {\sum\limits_{\Gamma J} {\sum\limits_{\Gamma 'J'} {{d_{\Gamma J}}{{d'}_{\Gamma 'J'}}{{\left( { - 1} \right)}^{J - M_J}}\left( {\begin{array}{*{20}{c}}
\begin{array}{l}
J\\
 - M_J
\end{array}&\begin{array}{l}
1\\
q
\end{array}&\begin{array}{l}
{J'}\\
{M'_{J'}}
\end{array}
\end{array}} \right)} } \left\langle {\Gamma J\left\| {{\textbf{P}^{\left( 1 \right)}}} \right\|\left. {\Gamma 'J'} \right\rangle } \right.} \right|} ^2}
\end{equation}
where $\lambda$ is the wavelength in $\AA$ and $\bf{P}^{(1)}$ is the electric dipole transition operator.
\subsubsection{MCDHF method}
According to the MCDHF method, the atomic state wave function (ASF) $\Psi (\Gamma JM_J)$ is a linear combination of a number of configuration state functions (CSFs) $\Phi_j (\gamma_j JM_J)$ with the same parity $P$, total angular momentum $J$ and its component along $z$ direction $M_J$
\begin{equation}
\Psi (\Gamma JM_J)=\sum_{j}^{N} {c_j} {\Phi_j(\gamma_jJM_J)},
\end{equation}
where $c_j$ stands for the mixing coefficient, $\gamma$ represents the other quantum numbers to uniquely define the state. The configuration state functions $\Phi_j(\gamma_jJM_J)$ are constructed as linear combinations of Slater determinants, each of which is a product of one-electron Dirac orbitals.

In the self-consistent field (SCF) procedure, the coefficients $c_j$ and the one-electron relativistic orbitals are optimized by solving the MCDHF equations, which are derived from the variational principle. The Breit interaction
\begin{equation}
\bm{B}_{ij} = - \frac{1}{2r_{ij}}[ {\bm{\alpha} _i} \cdot {\bm{\alpha} _j} + \frac{( {\bm{\alpha} _i} \cdot {r_{ij}})({\bm{\alpha} _j} \cdot {r_{ij}})} {r_{ij}^2}]
\end{equation}
and QED effects are included in the subsequent relativistic configuration interaction (RCI) calculation, where only the mixing coefficients are variable.
\subsection{Computational method}
In our calculations, we used the active space method to capture the electron correlation. The $1s^22s^22p^63s^23p^63d^{10}4s^24p^65s^2$ and $1s^22s^22p^63s^23p^63d^{10}4s^24p^65s5p$ configurations were treated as the reference configuration for the ground state ($5s^2$~$^1S_0$) and the excited states ($5s5p$~$^{1,3}P$), respectively, where the $5s$ and $5p$ electrons are the valence electrons and the others the core. The configuration expansions were generated by single (S) and double (D) excitations from the reference configuration to the active set. Starting with the ground state, the occupied spectroscopic orbitals in the reference configurations were optimized in the Dirac-Hartree-Fock (DHF) approximation and kept frozen in the following computations, and others in the active set as correlation orbitals. The valence-valence (labeled as VV) and core-valence (labeled as CV) correlations between the cores with n = 3, 4 and the valence electrons were systematically considered in the subsequent SCF calculation procedure. The SD excitations were restricted that at most one electron may be promoted from the core shells. The active sets were expanded as
\begin{equation}
\begin{split}
  n5  &=\{3s, 3p, 3d, 4s, 4p, 4d, 4f, 5s\}, \\
  n6  &=n5  + \{5p, 5d, 5f, 5g, 6s\}, \\
  n7  &=n6  + \{6p, 6d, 6f, 6g, 7s\}, \\
  n8  &=n7  + \{7p, 7d, 7f, 7g, 8s\}, \\
  n9  &=n8  + \{8p, 8d, 8f, 9s\}, \\
  n10 &=n9  + \{9p, 9d, 10s\}, \\
  n11 &=n10 + \{10p, 11s\}.
\end{split}
\end{equation}
The seven layers of virtual orbitals were added to make sure the convergence of the atomic parameters under investigation. Only the added orbitals in each layer of the active set were varied. Calculations for the excited states were performed in the same way except for the first layer of the active set $n$5 = \{3$s$, 3$p$, 3$d$, 4$s$, 4$p$, 4$d$, 4$f$, 5$s$, 5$p$\}.

To consider the core-core (labeled as CC) correlation of the n = 4 core shell, the CSFs by unrestricted SD excitations from the reference configuration to the active set of orbitals with n = 8 were generated. Furthermore, the higher-order electron correlations among the n = 4 and n = 5 shells were taken into account by the multi-reference (marked as MR) SD model. The final MR computational model contains the higher-order electron correlations as well as the VV, CV and CC correlations. In this step, the \{$4s^24p^65p^2$; $4s^24p^54d5s5p$\} and \{$4s^24p^64d5p$; $4s^24p^65s6p$; $4s^24p^65p6s$; $4s^24p^65s5p5d^2$\} configurations were added to the single reference configuration set. The configuration space was expanded by replacing one or two electrons in the reference configurations with ones in the active set $n$8. Finally, the Breit interaction and the QED correlations were evaluated. These calculations were performed in the RCI computation. We used the GRASP2K package \cite{Jonsson20132197} to accomplish our calculations.
\subsection{Numerical results and discussions}
\subsubsection{Excitation energies and rates of $^1S_0$~-~$^{1,3}P_1$ $E$1 transition}
Table \hyperref[1]{\uppercase\expandafter{\romannumeral1}} shows the excitation energies (in cm$^{-1}$) and the transition rates (in s$^{-1}$) of $^1S_0$~-~$^{1,3}P_1$ $E$1 transitions with different computational models. The transition rates in the Babushkin and the Coulomb gauges, corresponding to the nonrelativistic length and velocity gauges \cite{0022-3700-7-12-007}, are also displayed in this table. It can be seen from this table that the excitation energies and the transition rates converged very well when the virtual orbitals increased from $n$5 to $n$11. Comparing the results obtained from the CC model with those from $n$11, we found that the core-core correlation changes the atomic parameters considerably. For example, the $^1S_0$~-~$^3P_1$ transition energy decreases from 14538.74 cm$^{-1}$ to 13064.84 cm$^{-1}$. However, the higher-order correlations counteract the core-core effects. The effects of the Breit interaction and the QED corrections are tiny, and thus included in the uncertainty. Two different methods, based on the convergence trend of the atomic parameters and the consistence between the transition rates in two gauges \cite{PhysRevA.78.062505}, were used to estimate the uncertainties of the transition rates. The error bar of the $^1S_0$~-~$^1P_1$ $E$1 transition rate reaches 2\%, but 4\% for the rate of $^1S_0$~-~$^3P_1$ transition.

For comparison, other theoretical and experimental values are also presented in Table \hyperref[1]{\uppercase\expandafter{\romannumeral1}}.
As far as the excitation energy of the $^1S_0$~-~$^1P_1$ transition is concerned, our result agrees with those in
Ref.~\cite{PhysRevA.65.042503} obtained by using the configuration interaction (CI) and many-body perturbation theory (MBPT), in
Ref.~\cite{PhysRevA.75.014502} by the MCDF method and in Ref.~\cite{PhysRevA.88.022511} by the configuration interaction plus core polarization (CICP) method. For the excitation energy of the $^1S_0$~-~$^3P_1$ transition, our calculation result is 0.6\% lower than that obtained in Ref.~\cite{PhysRevA.64.012508,PorsevKozlovRakhlina} by the CI + MBPT method. Comparing with the experimental values from the National Institute of Standards and Technology (NIST) database \cite{Sansonetti}, we found that the relative uncertainties in the excitation energies of $^1S_0$~-~$^{1,3}P_1$ transitions are 0.9\% and 1.9\%, respectively.

For the $^1S_0$~-~$^1P_1$ $E1$ transition rate, our calculated results are in excellent agreement with ones obtained with the relativistic CI method \cite{0953-4075-36-17-306} and the latest CICP computational model \cite{PhysRevA.88.022511}. Compared with the experimental measurements, our calculations are consistent with the result of Parkinson \textit{et al.}~\cite{0022-3700-9-2-006} using the hook method (distortion of interferometric channel-spectra in the neighbourhood of absorption lines by anomalous dispersion) and the recent measurement of Nagel \textit{et al.}~\cite{PhysRevLett.94.083004} using photoassociative spectroscopy of $^{88}$Sr$_2$. The numbers in the round brackets are the uncertainties of their measurements. For the $^1S_0$~-~$^3P_1$ intercombination transition, our calculated transition rates are in good agreement with the MCDF result \cite{PhysRevA.75.014502}. Moreover, the calculated transition rates in our calculations are about 0.4\% larger than the measurements by observing the exponential decay of the florescence from the $^3P_1$ excited state \cite{ZphysD41}.
\begin{table}[h]\scriptsize
\caption{\label{1}Different computational models for the excitation energies $E$ (in cm$^{-1}$) and the transition rates (in s$^{-1}$) of $^1S_0$~-~$^{1,3}P_1$ (B: Babushkin gauge; C: Coulomb gauge). The numbers in square brackets are the expansion in base 10.}
\begin{ruledtabular}
\begin{tabular*}{8.6 cm}{ccccccc}
\multirow{2}{*}{Model} &      \multicolumn{3}{c}{$^1S_0$~-~$^3P_1$}   &        \multicolumn{3}{c}{$^1S_0$~-~$^1P_1$} \\\cline{2-7}
                       & $E_{13}$    &    B       &       C         & $E_{11}$ &     B            &      C     \\\hline
DHF                    &  8455.93    &   2.24[3]  &     4.32        & 23583.72 &    4.12[8]       &     1.90[8]    \\
n5                     & 15231.16    &   4.03[4]  &     5.83[4]     & 23086.27 &    2.44[8]       &     2.30[8]    \\
n6                     & 14745.84    &   5.05[4]  &     5.06[4]     & 21942.01 &    1.99[8]       &     1.81[8]    \\
n7                     & 14648.86    &   5.04[4]  &     4.28[4]     & 21861.95 &    1.96[8]       &     1.84[8]    \\
n8                     & 14568.70    &   5.00[4]  &     4.20[4]     & 21796.77 &    1.94[8]       &     1.86[8]    \\
n9                     & 14555.01    &   4.99[4]  &     4.11[4]     & 21760.25 &    1.91[8]       &     1.86[8]    \\
n10                    & 14548.28    &   4.95[4]  &     4.01[4]     & 21755.66 &    1.90[8]       &     1.86[8]    \\
n11                    & 14538.74    &   4.96[4]  &     3.95[4]     & 21746.22 &    1.91[8]       &     1.85[8]    \\
CC                     & 13064.84    &   1.92[4]  &     5.83[3]     & 23817.56 &    3.30[8]       &     2.13[8]    \\
MR                     & 14588.77    &   4.71[4]  &     4.46[4]     & 21904.02 &    1.96[8]       &     2.03[8]    \\
                                                              & \multicolumn{6}{c}{Other theories}\\
Porsev \textit{et al.} \cite{PhysRevA.64.012508,PorsevKozlovRakhlina}&14598   &5.35[4]&6.11[4]   &21621   & 1.92[8]    & 1.95[8]        \\
Savukov \textit{et al.} \cite{PhysRevA.65.042503}                    &15081   &5.55[4]&5.69[4]   &21981   & 1.89[8]    & 1.87[8]         \\
  Dzuba \textit{et al.} \cite{PhysRevA.68.022506}                    &14384   &       &          &22829   &            &             \\
Liu \textit{et al.} \cite{PhysRevA.75.014502}                    &14343.64&\multicolumn{2}{c}{4.52[4]}&21628.84&\multicolumn{2}{c}{1.89[8]}\\
Glowacki \textit{et al.} \cite{0953-4075-36-17-306}              &        &\multicolumn{2}{c}{3.54[4]}&        &\multicolumn{2}{c}{1.98[8]}\\
Vaeck \textit{et al.} \cite{0953-4075-24-2-006}                  &        &       &                   &        &\multicolumn{2}{c}{2.22[8]}\\
Safronova \textit{et al.} \cite{PhysRevA.87.012509}              &        &\multicolumn{2}{c}{5.16[4]}&        &\multicolumn{2}{c}{1.87[8]}\\
Cheng \textit{et al.} \cite{PhysRevA.88.022511}                  &14702.61&       &                   &21698.47&\multicolumn{2}{c}{1.93[8]}\\
                                                              & \multicolumn{6}{c}{Experiments}\\
NIST \cite{Sansonetti}                            &14317.51 & &                             &21698.45 &          &                    \\
Parkinson \textit{et al.} \cite{0022-3700-9-2-006}&         &\multicolumn{2}{c}{4.40(12)[4]}&         &\multicolumn{2}{c}{2.01(6)[8]} \\
Kelly \textit{et al.} \cite{Kelly}                &         & &                             &         &\multicolumn{2}{c}{2.14(5)[8]} \\
Husain \textit{et al.} \cite{F29858100087}        &         &\multicolumn{2}{c}{4.99(10)[4]}&         &   \\
Kelly \textit{et al.} \cite{PhysRevA.37.2354}     &         &\multicolumn{2}{c}{4.55(10)[4]}&         &    \\
Drozdowski \textit{et al.} \cite{ZphysD41}        &         &\multicolumn{2}{c}{4.69(11)[4]}&         &   \\
Nagel \textit{et al.} \cite{PhysRevLett.94.083004}&         & &                             &         &\multicolumn{2}{c}{1.92(1)[8]} \\
Yasuda \textit{et al.} \cite{PhysRevA.73.011403}  &         & &                             &         &\multicolumn{2}{c}{1.90(0.1)[8]}   \\
\end{tabular*}
\end{ruledtabular}
\end{table}
\subsubsection{Magnetic-field-induced transition}
For a bosonic Sr atom, the external magnetic field mixes the $^3P_1$ and $^1P_1$ states with $^3P_0$, and thus opens up a one-photon $E1$ transition channel from the $^3P_0$ state to the ground state. Since the magnetic field strength is weak and the energy separation between the states belonging to different configurations is large, the reference $^3P_0$ state can be approximately expressed as \cite{PhysRevA.88.013416}
\begin{equation}
\begin{aligned}
\left| 5s5p~^3P_0',M=0 \right\rangle = & \left| 5s5p~^3P_0,M=0 \right\rangle + \\
& \sum\limits_{s=1,3} {d_{s;J=1}} \left|5s5p~^sP_1,M=0 \right\rangle.
\end{aligned}
\end{equation}
Here, the state with prime describes the dominant component of the eigenvector. Isolated from other states, the ground state is given as
\begin{equation}
\left| 5s^2~^1S_0',M=0 \right\rangle = \left| 5s^2~^1S_0,M=0 \right\rangle.
\end{equation}

The expansion coefficient defined in Eq.(7) is simplified as
\begin{equation}
{d_s} = \frac{\langle {^sP_1} \|{\bm{N}^{(1)} + \Delta \bm{N}^{(1)}} \| {^3P_0}\rangle \textrm{B}}{E({^3P_0}) - E({^sP_1})}= {d^R_s} \textrm{B},
\end{equation}
where $d^R_s$ is the reduced mixing coefficient and the off-diagonal reduced magnetic interaction matrix element is obtained by using the HFSZEEMAN package \cite{Hfszeeman}.

Inserting Eqs.~(12) and (13) into Eq.~(8), we obtain the magnetic-field-induced the $5s5p$~$^3P_0$~-~$5s^2$~$^1S_0$ transition rate
\begin{equation}
\begin{aligned}
A_{MIT}(^1S_0,^3P_0) & =  \frac{2.02613 \times {{10}^{18}}}{3{\lambda ^3}} \times \\
& |\sum\limits_{s=1,3} {d^R_s} \langle {5s^2~^1S_0} \| {\bf P}^{(1)} \| {5s5p~^sP_1} \rangle \textrm{B}|^2.
\end{aligned}
\end{equation}

In Table \hyperref[2]{\uppercase\expandafter{\romannumeral2}} we present the off-diagonal reduced magnetic interaction matrix elements $W$ (in a.u.) and the reduced mixing coefficients $d_s^R$ (in T$^{-1}$) for various computational models. To obtain an accurate value of the MIT rate, we include both the $^1P_1$ and $^3P_1$ perturbations. The MIT rate (in s$^{-1}$) for the $^{88}$Sr atom in an external magnetic field with the strength B (in T) can be expressed as,
\begin{equation}
A_{MIT}({^1S_0},{^3P_0}) =A^R_{MIT}({^1S_0},{^3P_0})\cdot{\textrm{B}^2}= 0.198\cdot{\textrm{B}^2}.
\end{equation}
The unit of the reduced mixing coefficients $d^R_s$ is obtained by a conversion factor 2.353 $\times$ 10$^5$~a.u./T and the reduced transition rate $A^R_{MIT}$($^1S_0$,$^3P_0$) = 0.198~s$^{-1}$T$^{-2}$.
\begin{table}[h]
\caption{\label{2}Off-diagonal reduced magnetic interaction matrix elements $W$ (in a.u.) and reduced mixing coefficients $d_s^R$ (in T$^{-1}$) for various computational models. The numbers in square brackets are the expansion in base 10.}
\begin{ruledtabular}
\begin{tabular*}{8.6cm}{ccccc}
\multirow{2}{*}{Model}& \multicolumn{2}{c}{($^3P_1$,$^3P_0)$}&\multicolumn{2}{c}{($^1P_1$,$^3P_0$)}\\ \cline{2-5}
                     &      $W$    &   $d_3^R$             &     $W$       &  $d_1^R$\\ \hline
DHF                  &    0.40913  &  -2.090[-3]           & -0.40914[-2]  &  2.507[-7]\\
n5                   &    0.40899  &  -2.062[-3]           & -0.11304[-1]  &  1.313[-6]\\
n6                   &    0.40895  &  -2.011[-3]           & -0.12762[-1]  &  1.611[-6]\\
n7                   &    0.40896  &  -2.016[-3]           & -0.12614[-1]  &  1.589[-6]\\
n8                   &    0.40895  &  -1.998[-3]           & -0.12622[-1]  &  1.585[-6]\\
n9                   &    0.40895  &  -1.991[-3]           & -0.12624[-1]  &  1.594[-6]\\
n10                  &    0.40895  &  -1.991[-3]           & -0.12531[-1]  &  1.581[-6]\\
n11                  &    0.40895  &  -1.991[-3]           & -0.12507[-1]  &  1.577[-6]\\
CC                   &    0.40908  &  -2.077[-3]           & -0.76010[-2]  &  6.502[-7]\\
MR                   &    0.40898  &  -2.141[-3]           & -0.11295[-1]  &  1.407[-6]\\
\end{tabular*}
\end{ruledtabular}
\end{table}
\subsection{Comparison with the experimental measurements}
The line-width of the forbidden transition depends on the magnetic field strength. To obtain an ultra-narrow line-width, the static magnetic field strength should be extremely small \cite{Baillard:07,PhysRevLett.96.083001}. In this experiment, the external magnetic field strength is set to be below 5~mT.

In our previous experimental work, we measured the Rabi oscillation with different Rabi frequencies. The data were fitted with the function $P = a(1-cos(2\pi\Omega t)exp(-t/\tau))$ \cite{COL.30.100201,Tino2014} where $\tau$ is the decoherent time scale which remain constant. The Rabi frequency $\Omega$ of the MIT is defined as \cite{PhysRevLett.96.083001},
\begin{equation}
\Omega  = \frac{1}{{{\hbar ^2}}}\left[ {d_1^R\left\langle {{}^1{S_0}} \right.\left\| {{P^{\left( 1 \right)}}} \right\|\left. {{}^1{P_1}} \right\rangle  + d_3^R\left\langle {{}^1{S_0}} \right.\left\| {{P^{\left( 1 \right)}}} \right\|\left. {{}^3{P_1}} \right\rangle } \right]\left( {\mathord{\buildrel{\lower3pt\hbox{$\scriptscriptstyle\rightharpoonup$}}
\over E}  \cdot \mathord{\buildrel{\lower3pt\hbox{$\scriptscriptstyle\rightharpoonup$}}
\over B} } \right) = \alpha \sqrt I \left| B \right|\cos \theta.
\end{equation}
Here, $\langle {^1S_0} \|{P^{(1)}}\| {^1P_1}\rangle$ and $\langle {^1S_0} \|{P^{(1)}}\| {^3P_1}\rangle$ are the reduced matrix elements for the $^1S_0$~-~$^{1,3}P_1$ transitions. The values of the two matrix elements and the reduced mixing coefficients were accurately determined in previous calculations. The Rabi frequency is a function of the magnetic field strength $B$ as the intensity of 698-nm laser $I$ is kept constant. For a two-level atom, the excitation fraction $P$ is defined as
\begin{equation}
P = \frac{\Omega B \tau }{1 + {{{{\Omega ^2}\tau } \mathord{\left/ {\vphantom {{{\Omega ^2}\tau } {{A_{MIT}}}}} \right.
 \kern-\nulldelimiterspace} {{A_{MIT}}}}} }\left[ {d_1^R\left\langle {{}^1{S_0}} \right.\left\| {{P^{\left( 1 \right)}}} \right\|\left. {{}^1{P_1}} \right\rangle  + d_3^R\left\langle {{}^1{S_0}} \right.\left\| {{P^{\left( 1 \right)}}} \right\|\left. {{}^3{P_1}} \right\rangle } \right] \sin (\Omega \pi t).
\end{equation}
In this experiment, we used the $\pi$-pulse (t = 1/(2$\Omega$)) laser to excite atoms to the excited state and the intensity of 698-nm laser is 3.6~W/cm$^2$. In Fig.~3, we present the experimental measurements and the theoretical calculations of the excitation fraction. The solid squares are the weighted average of several measured excitation fractions with the magnetic field strength $B$ = 1.0, 1.2, 1.7, 2.2~mT and the error bars are the statistical uncertainty of our measurements. The red line is the calculated excitation fraction as a function of the magnetic field strength. As can be seen from this figure, our calculated results agree with the measurements when $B$ = 1.2, 1.7, 2.2~mT. The disagreement in the small magnetic field strength results from the low signal-to-noise ratio of the spectrum duo to the fluctuation of the cold atom number in the optical lattice. In practice, the 698-nm laser pulse cannot excite all of the atoms in the optical lattice to the $^3P_0$ state (about 60\% of the maximum excitation), and we assume that the excitation is substantially faster than the dominant loss mechanisms, for example, the frequency and amplitude fluctuation of the 813-nm laser and the vibration of the achromatic lens and the retro-reflecting mirror.
\begin{figure}
 \includegraphics{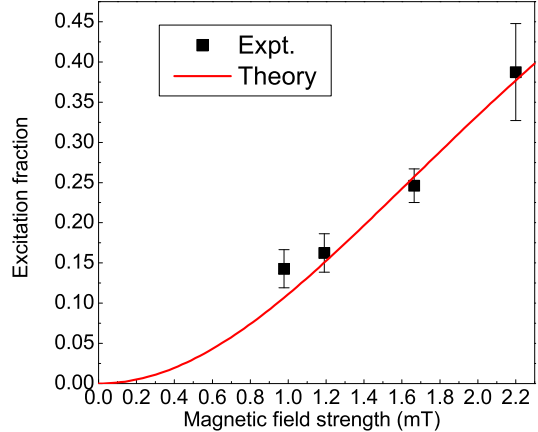}
 \caption{\label{5}The $5s5p$~$^3P_0$~-~$5s^2$~$^1S_0$ excitation fraction for different magnetic field strengths. The red line is the calculated excitation fraction as a function of magnetic field strength. The solid squares are the weighted average of several measured excitation fractions with the magnetic field strength $B$ = 1.0, 1.2, 1.7, 2.2~mT and the error bars are the statistical uncertainty of our measurements.}
\end{figure}
\section{CONCLUSION}
Actually, several research groups have reported the operation and characterization of the magnetic-field-induced the $nsnp$~$^3P_0$~-~$ns^2$~$^1S_0$ forbidden transition in bosonic atoms in the previous publications. However, the transition rate has never been determined in these works. In this work, we measured the excitation fraction of the $5s5p$~$^3P_0$~-~$5s^2$~$^1S_0$ transition using the normalized detection method under the circumstance of the different magnetic field strengths in the $^{88}$Sr ultra-cold atomic system. Meanwhile, we systematically analyzed the mechanism of this transition and accurately calculated the forbidden transition rate. Based on the calculations, we deduced the excitation fraction as a function of the magnetic field strength. The calculations are consistent with the experiments. The measurements indicated that our theoretical calculation model is reasonable. Moreover, the calculations also verified that the experimental method to determine the magnetic field strength effects on the lifetime of the metastable state we have developed is correct. This study should be helpful in more accurate evaluating the magnetic field effect on the energy level shift. Based on this, we can accurately estimate the magnetic field effect on the hyperfine energy level and the higher-order Zeeman shift of the $^{87}$Sr atom. Moreover, we can estimate the external magnetic field effects on the ultra-cold atoms collision.

\begin{acknowledgments}
This work was supported by National Natural Science Foundation of China (Grant No. 61127901, 11404025 and 91536106), the Strategic Priority Research Program of the Chinese Academy of Sciences (Grant No. XDB21030700), Space strontium optical atomic clock key technology research (Grant No. QYZDB-SSW-JSC004) and the China Postdoctoral Science Foundation (Grant No. 2014M560061).
\end{acknowledgments}

\bibliography{mylib}

\end{document}